\documentclass[twocolumn,showpacs,floatfix,superscriptaddress,amsmath,amssymb,aps,10pt,prl]{revtex4-2}

\usepackage{graphicx}
\usepackage{dcolumn}
\usepackage{amsmath}
\usepackage{bm}
\usepackage{appendix}
\usepackage{multirow}
\usepackage[bookmarks=true,colorlinks=true,urlcolor=blue,linkcolor=blue,citecolor=blue,breaklinks]{hyperref}
\usepackage[mathlines]{lineno}
\usepackage{bbold}
\usepackage{inconsolata} 
\usepackage[usenames,dvipsnames]{color} 
\usepackage{listings}
\usepackage{cancel}
\usepackage{soul}
\usepackage{xcolor}
\usepackage[normalem]{ulem}

\newcommand{\singapore}{Department of Physics, National University of Singapore, Singapore}
\newcommand{\majulab}{MajuLab, CNRS-UCA-SU-NUS-NTU International Joint Research Unit, Singapore}
\newcommand{\CQT}{Centre for Quantum Technologies, National University of Singapore, Singapore}
\newcommand{\como}{Center for Nonlinear and Complex Systems, Dipartimento
di Scienza e Alta Tecnologia, Universit\`a degli Studi dell'Insubria,
via Valleggio 11, 22100 Como, Italy}
\newcommand{\infn}{Istituto Nazionale di Fisica Nucleare, Sezione di Milano,
via Celoria 16, 20133 Milano, Italy}
\newcommand{\brazil}{International Institute of Physics, Federal University
of Rio Grande do Norte, Campus Universit\'ario - Lagoa Nova, CP. 1613,
Natal, Rio Grande Do Norte 59078-970, Brazil}
\newcommand{\xiamen}{Department of Physics and Fujian Provincial Key
Laboratory of Low Dimensional Condensed Matter Physics, Xiamen University,
Xiamen 361005, Fujian, China}
\newcommand{\lanzhou}{Lanzhou Center for Theoretical Physics, Lanzhou
University, Lanzhou 730000, Gansu, China}

\begin{document}

\preprint{APS/123-QED}

\title{Quantum vs Classical Thermal Transport at Low Temperatures}

\author{Zhixing Zou}
\affiliation{\singapore}
\author{Jiangbin Gong}
\affiliation{\singapore}
\affiliation{\majulab}
\affiliation{\CQT}
\author{Jiao Wang}
\affiliation{\xiamen}
\affiliation{\lanzhou}
\author{Giulio Casati}
\affiliation{\como}
\affiliation{\brazil}
\author{Giuliano Benenti}
\affiliation{\como}
\affiliation{\infn}

\date{\today}

\begin{abstract}
This work aims to understand how quantum mechanics affects heat transport at low temperatures.
In the classical setting, 
by considering a simple paradigmatic model, our simulations
reveal the emergence of Negative Differential Thermal Resistance (NDTR): paradoxically, increasing the temperature bias by lowering the cold bath temperature reduces the steady-state heat current. In sharp contrast, the quantum version of the model, treated via a Lindblad master equation, exhibits no NDTR: the heat current increases monotonically with thermal bias. This marked divergence highlights the fundamental role of quantum effects in low-temperature thermal transport and underscores the need to reconsider classical predictions when designing and optimizing nanoscale thermal devices.

\end{abstract}

\maketitle

\textit{Introduction}
Understanding the microscopic origins of macroscopic thermodynamic behavior is a central goal of statistical physics. Classical physics fails to 
reproduce certain thermodynamic quantities — most notably the heat capacity of materials at low temperatures \cite{einstein1907,debye1912,kittel} — a deficiency that was one of the key motivations for the development of quantum theory. 
A similar need to distinguish between classical and quantum predictions emerges in the context of nonequilibrium thermodynamics, particularly in the study of heat transport. This issue is driven both by fundamental scientific interest and by the rapid advancement of solid-state nanodevices, where thermal properties at low temperatures play a critical role \cite{Pekola2021}. Indeed, the ability to control and exploit the thermal transport properties of a system can result in devices such as thermal switches \cite{Wehmeyer2017}, thermal amplifiers \cite{Zhong2012}, thermal logic gates \cite{Wang2007}, and thermal memories \cite{Wang2008}, which are at the core of a broad range of potential applications \cite{Li2012,Pekola2021}. 

The implementation of the above-mentioned thermal devices relies on the realization of the thermal transistor \cite{Li2006,Joulain2016,Sood2018,Wang2020,Li2023,Castelli2023,Lim2024}, 
which in turn is 
grounded in the phenomenon of Negative Differential Thermal Resistance (NDTR),  \cite{Li2004a,Li2006,Segal2006,Li2006,Wang2007,Yang2007,ChungLo2008,Zhong2009,Shao2009,He2009,He2010,Pereira2010,Zhong2011,Fu2015,Chan2014,Mendonca2015}
recently observed in classical hard-core gas models \cite{Luo2019,Luo2022}. 
Classical NDTR results so far have been obtained using the traditional Maxwell bath model \cite{Lebowitz1978}, which assumes that particles thermalize instantaneously upon interaction with the bath. However, since the NDTR mechanism proposed in these studies involves cooling one of the baths to very low temperatures, the absence of a finite-rate relaxation process in the bath model may significantly affect the nonequilibrium steady state and, consequently, the resulting heat flux.
Moreover, in the zero-temperature limit, a Maxwell bath istantaneously freezes colliding particles, an artificial feature in conflict with the third law of thermodynamics. 
These considerations give rise to fundamental questions: Is NDTR merely an artifact of the bath model, or does it reflect genuine features of classical dynamics? Furthermore, if the effect indeed originates from classical behavior, does it persist in a quantum mechanical framework? In other words: how does the heat current depend on the thermal gradient in the quantum regime at low temperatures?

In this work, we investigate both classical and quantum heat transport in a paradigmatic nonlinear system: a single particle confined within a one-dimensional channel by a square-well potential, with the ends of the channel coupled to thermal reservoirs at different temperatures. This minimal model captures
the fundamental features of heat transport at low temperatures
and provides a direct basis for comparing
classical and quantum behavior \footnote{This model represents the simplest instance of billiard-type gas models, which serve as paradigms in the study of classical nonequilibrium dynamics due to their conceptual simplicity and dynamical richness \cite{Lebowitz1978, Dettmann2000, Li2004, Benenti2013, Wang2017, Luo2018, Luo2019, Wang2020, Luo2022, Zou2024}.
In contrast, studies of quantum thermal transport have predominantly focused on spin chain and tight-binding systems \cite{Balachandran2018, Balachandran2019, Landi2022}. This preference likely stems from the intrinsic difficulty of modeling collisional interactions between gas particles and thermal baths in the quantum regime.}.
In the classical regime, we perform molecular dynamics simulations, and, to ensure consistency with the third law of thermodynamics, we incorporate relaxation dynamics into the bath model. Remarkably, even under these thermodynamically consistent conditions, NDTR  still emerges, indicating that this phenomenon
is an intrinsic feature of classical dynamics.
This naturally raises the question of whether such classical effect truly reflects physical reality, since low-temperature behavior is ultimately governed by quantum mechanics. We therefore extend our analysis to the quantum regime, modeling the system’s dynamics using a Lindblad master equation carefully 
designed to reproduce
classical thermal transport in the high-temperature limit.
In stark contrast to the classical case, the quantum model 
exhibits no NDTR. 
These findings demonstrate that classical models can significantly mispredict thermal transport at low temperatures, highlighting the crucial role of quantum effects in this regime.

\begin{figure}
    \centering
    \includegraphics[width=1.0\linewidth]{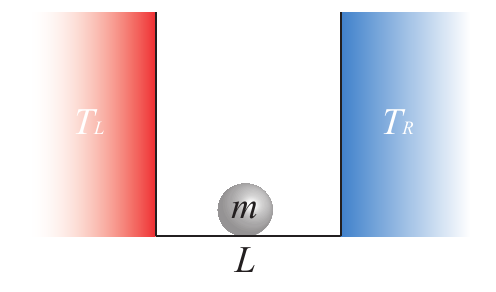}
    \caption{Schematic plot of the one-dimensional infinite square well system for heat transport. Each end of the system is connected to a thermal bath held at a different temperature.}
    \label{fig:fig1}
\end{figure}

\textit{Classical modeling and results.}
As depicted in Fig.~\ref{fig:fig1}, we consider
a particle of mass $m$ confined within a one-dimensional channel of length $L$ 
by a potential well. 
The system is described by the Hamiltonian 
\begin{equation}
    H = \frac{1}{2m} p^2 + V(x),
    \quad
    V(x) = \begin{cases}
        0, & 0<x<L,\\
        \infty, & {\rm otherwise},
    \end{cases}
\end{equation}
with each end of the channel connected to a thermal bath at different temperatures, 
$T_L$ and $T_R$.
Due to the interaction between the particle and the two baths, the system 
evolves toward a nonequilibrium steady state in the long-time limit, characterized by a  thermal flux.


In classical simulations, the bath is typically modeled as a Maxwell reservoir \cite{Lebowitz1978}: When the particle collides with bath $\alpha$ $(\alpha=L, R)$,
it is reflected with a random velocity $v$ sampled from
\begin{equation}
    f_\alpha (v)=\frac{m |v| }{k_BT_\alpha}\exp\left(-\frac{m v^2}{2k_BT_\alpha}\right),
\end{equation}
where $k_B$ is the Boltzmann constant.

According to the standard definition, the classical thermal flux $J^c$ from the left bath to the right bath in our system is given by
\begin{equation}\label{eq:cl_flux}
    J^c = \frac{E_{L\to R} - E_{R\to L}}{t_{L\to R} + t_{R\to L}},
\end{equation}
where $E_{L\to R}$ ($E_{R\to L}$) and $t_{L\to R}$ ($t_{R\to L}$) represent, respectively, the average energy the particle transfers and the average time it takes from the left (right) bath to the right (left) bath. 

For a particle coupled to Maxwell baths, the thermal flux can be calculated analytically.
The energy transferred from the left (right) bath during a single collision is
$E_{L\to R,R\to L} = \int_0^\infty \frac{1}{2}m v^2 f_{L,R}(v){\rm d} v = k_B T_{L,R}$ 
and the average time between successive collisions with the baths is
$ t_{L\to R,R\to L} = \int_0^\infty \frac{L}{|v|} f_{L,R}(v){\rm d} v = L\sqrt{\frac{m\pi}{2k_B T_{L,R} }}$. 
Substituting these expressions into Eq.~\eqref{eq:cl_flux}, we obtain the classical thermal flux for the Maxwell bath:
\begin{equation}\label{eq:maxwellflux}
    J^c = \frac{1}{L}\sqrt{\frac{2k_B}{m\pi}}\sqrt{T_LT_R}(\sqrt{T_L}-\sqrt{T_R}).
\end{equation}

The Maxwell baths approach assumes that the particle 
instantaneously reaches thermal equilibrium 
upon interacting with the bath.
However, this assumption 
becomes problematic, particularly at low temperatures. Consider the extreme case of a bath at absolute zero ($T=0$): 
under the Maxwell model, a collision with the bath would immediately reduce the particle's velocity to zero, effectively freezing it at the boundary.
This behavior contradicts the third law of thermodynamics which states that absolute zero cannot be reached in finite time. Consequentely, relaxation process cannot be neglected and must be properly accounted for.

To avoid such an unphysical scenario, we need to incorporate relaxation effects into the Maxwell bath. Relaxation is often described as a Markovian process, 
which can be conveniently modeled using the Markov Chain Monte Carlo (MCMC) method 
\cite{Heermann1986}. In this approach, a Markovian transition matrix $\mathcal{T}(v\to v')$ with steady distribution $f_\alpha (v)$ is used. After a 
sufficiently large number of Monte Carlo steps, the Markov chain converges to 
this distribution, thereby generating the target ensemble.
This naturally suggests 
interpreting the Markovian process in MCMC as a model that mimics the system’s relaxation dynamics. We refer to this scheme
as the MCMC Maxwell bath model [see Supplemental Material (SM) for details] 

In Fig.~\ref{fig:fig2}, the comparative results between traditional Maxwell baths and those implemented via MCMC are illustrated.
When the cold bath temperature $T_R$ is high, both models agree well, as the particle reflected by the MCMC Maxwell bath 
effectively relaxes
to thermal equilibrium (see SM). However, as $T_R$ decreases, deviations emerge
due to incomplete thermalization \footnote{In the MCMC Maxwell bath, incomplete relaxation leads to an effective temperature
$T_{eff}$ of the particle reflected from the cold bath that is higher than the prescribed bath temperature $T_{R}$. When $T_{eff}$ only slightly exceeds $T_{R}$, this enhancement results in a larger heat flux compared to the standard Maxwell bath, owing to the NDTR effect. At the special point where $T_{eff} = (\sqrt{T_{L}} - \sqrt{T_{R}})^2$, 
both models yield identical results, as can be seen from Eq.~(\ref{eq:maxwellflux}).
As $T_R$ decreases further, $T_{eff} > (\sqrt{T_{L}} - \sqrt{T_{R}})^2$ and
the heat flux from the MCMC Maxwell bath becomes smaller than that of the standard Maxwell bath.}.
In the limit $T_R\to 0$, the thermal flux in both models approaches zero, but for fundamentally different reasons. In the traditional Maxwell bath, 
this behavior stems
unphysically from 
a violation of the third law of thermodynamics: the particle instantaneously equilibrates with the cold bath, causing the time interval 
$t_{R\to L} $ in Eq.~\eqref{eq:cl_flux} to diverge. 
By contrast, in the MCMC Maxwell bath 
the vanishing flux 
results from the
lack of relaxation at the cold bath, which causes the energy difference $E_{L\to R}-E_{R\to L}$ in Eq.~\eqref{eq:cl_flux} to approach $0$. Clearly, both models exhibit the NDTR effect, as 
evidenced by the positive slopes of the $J^{c}$ vs $T_R$ plot in Fig.~2 over a wide range of $T_R$. 

\begin{figure}
    \centering
    \includegraphics[width=0.95\linewidth]{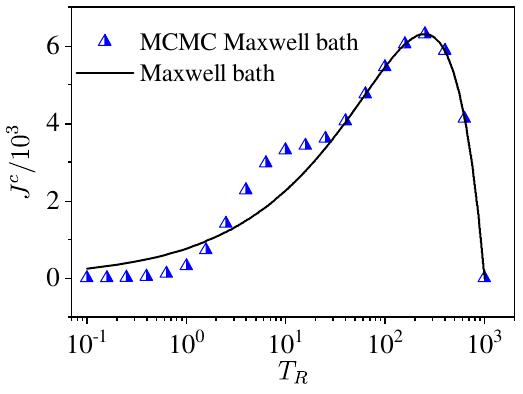}
    \caption{Dependence of classical thermal flux with cold bath temperature $T_R$. The line is for result obtained with Maxwell baths [Eq.~\eqref{eq:maxwellflux}], and the triangles are for MCMC Maxwell baths. In all simulations, we set $T_{L}=10^3$, $m=k_B = 1$ and, unless otherwise specified, $L=1$. }
    \label{fig:fig2}
\end{figure}


\textit{Quantum modeling and results.}
To simulate the quantum dynamics, we model the effect of the thermal baths using a Lindblad master equation \cite{Breuer2002}, consistent with the Markovian framework adopted in the classical case above:
\begin{equation}\label{eq:GME}
  \frac{{\rm d} \hat{\rho}}{{\rm d} t} = -\frac{i}{\hbar}\left\lbrack \hat{H},\hat{\rho}\right\rbrack + \mathcal{D}_L(\hat{\rho}) +  \mathcal{D}_R(\hat{\rho})\equiv \mathcal{L}(\hat{\rho}),
\end{equation}
where $\hat{\rho}$ is the density operator of the system, $\hbar$ is the 
Planck constant, and $\mathcal{D}_\alpha(\hat{\rho})$ 
is the dissipator term due to the interaction with bath $\alpha$. 
The general expression of the dissipator term is 
given by \cite{Breuer2002}
\begin{equation}
    \begin{split}
        \mathcal{D}_{\alpha}(\hat{\rho}) = & \sum_{\omega >0}J(\omega)\lbrace \lbrack 1+n_\alpha(\omega)\rbrack\lbrack\hat{A}_\alpha(\omega)\hat{\rho}\hat{A}_\alpha^\dagger(\omega) - \\ &~~~~~~~~1/2(\hat{A}_\alpha^\dagger(\omega)\hat{A}_\alpha(\omega)\hat{\rho} + \hat{\rho}\hat{A}_\alpha^\dagger(\omega)\hat{A}_\alpha(\omega)) \rbrack + \\
        &~~~~~~~~n_\alpha(\omega)\lbrack\hat{A}_\alpha^\dagger(\omega)\hat{\rho}\hat{A}_\alpha(\omega) - \\
        &~~~~~~~~1/2(\hat{A}_\alpha(\omega)\hat{A}_\alpha^\dagger(\omega)\hat{\rho} + \hat{\rho}\hat{A}_\alpha(\omega)\hat{A}_\alpha^\dagger(\omega)) \rbrack\rbrace.
    \end{split}
\end{equation}
Here, $\hbar \omega = \epsilon_j - \epsilon_i$ is the energy difference of two eigenstates, $|\epsilon_j\rangle$ and $|\epsilon_i\rangle$ of $H$, $J(\omega)$ is the bath spectral function, $n_\alpha(\omega) = \lbrack\exp(\hbar\omega/k_BT_\alpha)-1\rbrack^{-1} $ is the Bose-Einstein distribution characterizing the thermal state of bosonic baths,  and the Lindblad operator $\hat{A}_\alpha(\omega)$ takes the form 
\begin{equation}\label{eq:Lind_Oper}
    \hat{A}_\alpha(\omega) = \sum_{i,j} |\epsilon_i\rangle\langle \epsilon_i|\hat{P}_\alpha |\epsilon_j\rangle\langle \epsilon_j|\delta_{\epsilon_j - \epsilon_i-\hbar\omega},
\end{equation}
 which describes the transition induced by the bath through the system operator $\hat{P}_\alpha$. 
In our model, 
we write $\hat{P}_L$ and $\hat{P}_R$ as
 \begin{equation}
 \label{eq:int_operator}
     \hat{P}_L \propto \int_0^\delta |x\rangle\langle x| {\rm d}x,
     \quad
     \hat{P}_R \propto \int_{L-\delta}^L |x\rangle\langle x| {\rm d}x,
\end{equation}
where $\delta$ is a small parameter representing the spatial range at the boundaries over which the system interacts with the baths. The $\delta$-dependent prefactors of $\hat{P}_L$ and $\hat{P}_R$ in Eq.~(\ref{eq:int_operator}) are determined by requiring well-behaved matrix elements of operarors  $\hat{A}_{\alpha}(\omega)$ in the limit of $\delta\rightarrow 0$
[see SM for the derivation and the explicit form of the operators $\hat{A}_\alpha(\omega)$].


The quantum thermal flux can be calculated using the continuity equation for energy: 
\begin{equation}
    \frac{{\rm d}}{{\rm d}t} \mathrm{Tr}\lbrack \hat{\rho}\hat{H}\rbrack = J_L - J_R,
\end{equation}
where 
\begin{equation}
    J_L = \mathrm{Tr} \lbrack\mathcal{D}_L(\hat{\rho})\hat{H} \rbrack\quad{\rm and} \quad J_R = -\mathrm{Tr} \lbrack\mathcal{D}_R(\hat{\rho})\hat{H}\rbrack
\end{equation}
denote the thermal flux from the left and right bath to the system, respectively.  In the steady state $\hat{\rho}_s$ the quantum thermal flux $J^q$ from the left bath to 
the right baths is given by
\begin{equation}
    J^q = J_L = -J_R. 
\end{equation}



In the quantum model, the steady state is calculated from the stationary solution of equation $\mathcal{L}(\hat{\rho})=0$ \cite{Landi2022}. We solve the equation by finding the eigenvector of the superoperator $\mathcal{L}$ corresponding to the zero eigenvalue. Although the dimension of Hilbert space is infinite, 
it is reasonable to truncate it 
to $N_{\rm tru}$ energy levels, ensuring that the condition $\exp(-(\epsilon_{N_{\rm tru} }-\epsilon_{1}) /k_B T_{\rm max})\ll 1$  holds ($T_{\rm max}$ is the maximum temperature in our simulations).  We set $N_{\rm tru} = 100$, $T_{\rm max}=1000 $, and $\hbar=1$ to guarantee that this condition is met.
    
Note that in the infinite square well system, there is a finite minimum energy gap given by $\hbar\omega_{\rm min} = \epsilon_{2}-\epsilon_{1}$. 
Since the spectral function $J(\omega)$ affects the dynamics only for frequencies $\omega>\omega_{\rm min}$, 
we are free to adopt the form $J(\omega) = \gamma \omega_c^{1-\eta}  \omega^\eta$ within the actual frequency range $\omega>\omega_{\rm min}$, even for negative $\eta$.  Indeed, we still always  assume $J(\omega)\to 0$ as
$\omega\to 0$ so that $J(\omega)$ remains a physical spectral function \cite{Breuer2002}.  

\begin{figure}
    \centering
    \includegraphics[width=1.0\linewidth]{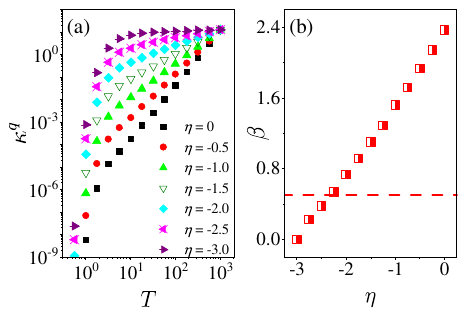}
    \caption{(a) Temperature dependence of thermal conductivity for different bath spectral functions $J(\omega)=\gamma \omega_c^{1-\eta}  \omega^\eta$, where $\gamma$ is always set through the relation $\kappa^q(T_{\rm max}) = \kappa^c(T_{\rm max})$. For $\eta$ = 0, -0.5, -1.0, -1.5, -2.0, -2.5, and -3.0, the corresponding values of $\gamma$ are $8.036\times 10^{-7}$, $3.755\times 10^{-5}$, $1.477\times 10^{-3}$, $4.691\times 10^{-2}$, $1.160$, $22.499$, and $357.057$, respectively.  Note that for all values of $\omega$ corresponding to transitions between energy eigenstates, i.e., $\hbar \omega = \epsilon_j - \epsilon_i$, the condition that the relaxation time is much shorter than the time scale of the intrinsic evolution of the system \cite{Breuer2002}, i.e. $\gamma J(\omega) << \omega$, is satisfied. (b) Relation between the exponent $\beta$ and the exponent $\eta$. The dashed line is for $\beta = \beta_{\rm cl} $. Hereafter, we set $\omega_c=1$.}
    \label{fig:fig3}
\end{figure}

To proceed with our quantum model analysis, it is necessary 
to determine the strength $\gamma$ in the spectral function $J(\omega)$. 
To achieve this, we enforce quantum-classical correspondence by matching the thermal conductivity at the highest temperature, i.e., 
$\kappa^q(T_{\rm max}) = \kappa^c(T_{\rm max})$. Figure~\ref{fig:fig3}(a) shows the temperature dependence of the thermal conductivity for different spectral functions of the form $J(\omega) = \gamma\omega_c^{1-\eta} \omega^\eta$ 
for $\omega\ge \omega_{\rm min}$. In the high temperature regime, the quantum thermal conductivity follows a scaling behavior $\kappa^q\propto T^\beta$, where $\beta$ is an $\eta$-dependent exponent. Notably, this temperature-dependent scaling has a parallel in classical systems. For a classical system at high temperature that can be well modeled with Maxwell baths, 
consider the case where $T_L = T+\frac{\Delta}{2}$, $T_R= T - \frac{\Delta}{2}$. As the temperature difference $\Delta$ between the two baths approaches zero, Eq.~\eqref{eq:maxwellflux} 
predicts that the classical thermal conductivity also follows a power-law scaling given by \footnote{Note that our definition deviates from the standard one commonly used in systems with finite particle \emph{density}: we normalize the thermal flux by the temperature difference, that is, we divide the thermal flux by the temperature difference, $\Delta$, rather than by the temperature gradient,
$\Delta/T$.}
\begin{equation}\label{eq:cl_kappa}
    \kappa^c = \lim_{\Delta\to 0} \frac{J^c}{\Delta} =\frac{1}{L}\sqrt{\frac{k_B T}{2m\pi}} \propto T^{1/2}.
\end{equation}
To fully determine the bath spectral function $J(\omega)$, we impose an additional restriction:
the quantum system must reproduce the same temperature-dependent behavior of the thermal conductivity as observed in the classical system in the high-temperature regime,
i.e., $\beta = \beta_{\text{cl}}=\frac{1}{2}$. From the relation between $\beta$ and $\eta$ (shown in Fig.~\ref{fig:fig3} (b)), we can extract an approximation for the value of $\eta$ that satisfies this condition, yielding $\eta \approx -2.25$.

\begin{figure}
    \centering
    \includegraphics[width=0.95\linewidth]{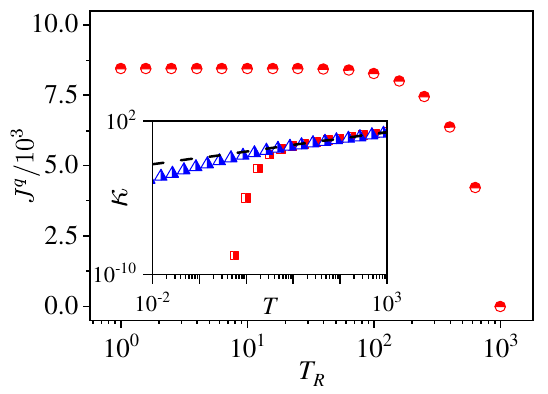}
    \caption{Dependence of quantum thermal flux with cold bath temperature $T_R$, with $T_L=10^3$.  The inset shows the temperature dependence of $\kappa$. Red squares are for quantum results, blue triangles are for classical results with MCMC Maxwell bath, and the dashed line is for the result of classical system with Maxwell baths [Eq.~\eqref{eq:cl_kappa}].}
    \label{fig:fig4}
\end{figure}

After determining the full bath spectral function, we analyze 
both the relaxation dynamics
(see SM) and thermal transport properties of the open square well model. 
The dependence of the quantum thermal flux $J^q$ on the cold bath
temperature $T_R$ is shown in Fig.~\ref{fig:fig4}. 
Notably no NDTR is observed: the plotted curve of $J_{q}$ vs $T_R$ has no positive slope over the entire range of $T_R$. 
To account for the absence of quantum NDTR observed here, we propose the following physical mechanism.
In the classical case, thermodynamically consistent NDTR arises from the vanishing relaxation rate of the cold reservoir in the zero-temperature limit, which suppresses energy exchange with that reservoir and ultimately leads to a breakdown of heat transport.
In contrast, the wavelike nature of the quantum particle enables continuous interaction with both thermal reservoirs, even at very low temperatures. Consequently, the classical mechanism responsible for
NDTR is no longer operative in the quantum regime, and hence NDTR does not emerge.

To further validate our approach, 
we present in the inset the temperature dependence of the thermal conductivity, defined within the linear response regime. At high temperatures, the quantum system indeed reproduces the same temperature scaling of thermal conductivity as its classical counterpart.
In the low-temperature regime, both the quantum results and those from the MCMC Maxwell bath deviate from the scaling law
(\ref{eq:cl_kappa}), 
with the quantum case exhibiting 
a pronounced super-exponential decay in thermal conductivity.
This quantum deviation from the classical thermal transport behavior indicates that the discreteness of the energy spectrum hinders bath-induced energy transitions at temperatures below the smallest energy gap between eigenstates, i.e., when $k_B T\ll \epsilon_2-\epsilon_1$.

Finally, we comment on the quantum-to-classical transition in our model. By applying the scaling $\gamma \to \gamma L^{-0.5}$, 
the quantum system exhibits a heat flux with the same 
$L$-dependence as in the classical case
(see simulation results in SM). 
This correspondence can be understood through the following scaling analysis.
To keep the Bose-Einstein distributions $n_\alpha((\epsilon_j - \epsilon_i)/\hbar)$ in Eq.~\eqref{eq:GME} invariant under changes in the system size $L$, the temperature must scale as $T/L^2$. With this scaling, the quantum heat flux becomes:
$J^q(T_L/L^2,T_R/L^2)\sim (\gamma L^{-0.5})(L^{-2})^{-2.25}(L^{-3})^2L^{-2}\sim L^{-4}$,
where the second term comes from the spectral density $J(\omega)$, the third from the Lindblad operators, and the last from the temperature difference between the baths.
In the classical case, using Eq.~\eqref{eq:maxwellflux}, the heat flux under the same temperature scaling also follows $J^c(T_L/L^2, T_R/L^2) \sim L^{-4}$. 
Therefore, the quantum and classical heat fluxes share the same $L$-dependence,
provided $\gamma$ scales as $L^{-0.5}$.
A natural consequence of this scaling behavior is that increasing $T_{\max}$ at fixed $L$ is equivalent to increasing $L$ at fixed $T_{\max}$—that is, both operations drive the quantum system toward the classical limit.

\textit{Conclusions and discussion.}
In summary, we have investigated thermal transport in a paradigmatic nonlinear system. 
On the classical side, using MCMC Maxwell baths to incorporate temperature-dependent relaxation effects, we observed the NDTR effect.
On the quantum side, we modeled the system using a Lindblad master equation to describe 
interactions with thermal baths. By appropriately designing the bath spectral function, 
the quantum model accurately reproduces the classical temperature dependence of thermal conductivity in the high-temperature regime. However, in stark contrast to the classical case, no evidence of NDTR is observed at low temperatures. These findings 
provide a foundational step toward understanding low-temperature thermal transport 
and may have 
practical implications for the design of nanoscale thermal devices.
From a theoretical perspective, extending this analysis beyond phenomenological open-system models to fully microscopic descriptions—-both classical and quantum--remains an important and challenging direction for future research.

Beyond purely thermal transport, billiard-like gas models provide a versatile platform for exploring coupled transport phenomena, such as thermoelectric effects \cite{Benenti2013,Luo2018}. Extending the present framework to investigate coupled transport in both classical 
and quantum open systems represents a promising 
avenue for future research. Such studies could 
yield valuable insights for the design of nanoscale energy conversion technologies \cite{Benenti2017} and quantum thermoelectric devices.

\textit{Acknowledgments.}  J.G. acknowledges support by the National Research Foundation, Singapore through the National Quantum Office, hosted in A*STAR, under its Centre for Quantum Technologies Funding Initiative (S24Q2d0009). J.W. is supported by the Natural Science Foundation of China (Grant No. 12475038) and the National Key R\&D Program of China (Grant No. 2023YFA1407100) and G.B. acknowledges support from INFN through the project QUANTUM. 
\bibliography{apssamp}

\newpage

\appendix

\begin{widetext}
\section*{Supplemental Material for ``Quantum vs Classical Thermal Transport at Low Temperatures"}

\subsection*{Details of the relaxation process in the MCMC Maxwell bath model}\label{sec:AppA}

First, we demonstrate that the MCMC Maxwell bath accurately captures the relaxation process. In our computations, we employ the Metropolis–Hastings algorithm  
\cite{Heermann1986} to generate the distribution $f_\alpha$. In this algorithm, the Markovian transition matrix is implicitly constructed as $\mathcal{T}(v\to v') = w(v'|v)A(v\to v')$, where $w(v'|v)$ is a proposed conditional distribution and $A(v\to v')=\min\left(1,f_\alpha(v')w(v|v')/f_\alpha(v)w(v'|v)\right)$ is the acceptance probability. In our computation to adjust the relaxation with temperature we set $w(v'|v)$ as a uniform distribution over the interval $(v-\sqrt{T}, v+\sqrt{T})$, i.e.,
$w(v'|v)= 0.5/\sqrt{T}$ for $v'\in (v-\sqrt{T},v+\sqrt{T})$. We adopt this form because the velocity of a bath particle is on the order of $\sqrt{T}$, so the change in particle velocity resulting from a single collision with one bath particle can likely be approximated as being of the same order. Additionally, we set a temperature-dependent Monte-Carlo step, which accounts for how many bath particles the system particle collides with when hitting the bath at the boundary. Physically, this is proportional to the number of bath particles hitting the boundary per unit time, which is of the order $\sqrt{T}$. Therefore, we take the number of Monte Carlo steps as $N_{MC} = c\sqrt{T}$, and first show results for $c=10$. Different choices of $c$ do not change our result qualitatively. 

We characterize the relaxation process by measuring the average energy after $N_c$ collisions with the baths, both held at the same temperature, $T_L=T_R=T$. 
Figure~\ref{fig:fig5}(a) shows the relaxation behavior of the particle coupled to MCMC Maxwell baths at various temperatures.  The initial velocity distribution is far from equilibrium, following a uniform distribution over the interval $(0, 2)$. Through continuous collisions with the bath, the particle gradually approach thermal equilibrium. Clearly, the relaxation process is temperature-dependent: fewer collisions are required for the particle to reach thermal equilibrium at higher bath temperatures, while more collisions are needed at lower temperatures.
To quantify this relaxation behavior, we extract the minimum number of collisions $N_c^*$
required for the particle to reach equilibrium at different bath temperatures. The results shown in Fig.~\ref{fig:fig5} (b) reveal that as the bath temperature approaches zero, $N_c^*$ diverges as $N_c^*\propto T^{-0.9} $, 
consistently with the third law of thermodynamics.

\begin{figure}[t]
    \centering
    \includegraphics[width=0.6\linewidth]{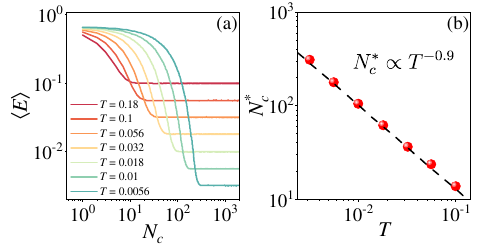}
    \caption{(a) Dependence of average energy of the particle on the number of collisions. The temperature decreases gradually from top to bottom. (b) Temperature dependence of the minimum number of collisions to reach thermal equilibrium. Dashed line is the power law fitting.}
    \label{fig:fig5}
\end{figure}



\subsection*{Expression  of the the dissipator terms}\label{sec:AppB}
For the infinite square well system, it's easy to solve the eigenvalue equation $\hat{H}|\epsilon_n\rangle = \epsilon_n |\epsilon_n\rangle$ and obtain its eigenvalues and eigenvectors as
\begin{equation}\label{eq:eigen}
    \epsilon_n = \frac{n^2\pi^2\hbar^2}{2mL^2},\quad |\epsilon_n\rangle = \sqrt{\frac{2}{L}}\int_0^L \sin\left(\frac{n\pi x}{L}\right) |x\rangle \mathrm{d} x,
\end{equation}
where $|x\rangle$ is the eigenfunction of position operator $\hat{x} | x \rangle = x | x \rangle$.

By substituting Eqs. \eqref{eq:eigen} and \eqref{eq:int_operator} into Eq.~\eqref{eq:Lind_Oper}, the Lindblad operators for the left and the right bath, $\hat{A}_L(\omega)$ and $\hat{A}_R(\omega)$, can be expressed as follows:
\begin{equation}
\begin{split}
    \hat{A}_L(\omega) &\propto \frac{1}{\pi} \sum_{i,j}\left(\frac{\sin\left(\frac{\pi\delta(i-j)}{L}\right)}{i-j} - \frac{\sin\left(\frac{\pi\delta(i+j)}{L}\right)}{i+j}\right)|\epsilon_i\rangle\langle\epsilon_j|\delta_{\epsilon_j - \epsilon_i-\hbar\omega},\\
    \hat{A}_R(\omega) &\propto \frac{1}{\pi} \sum_{i,j}\left(\frac{\sin\left(\frac{\pi(L-\delta)(i-j)}{L}\right)}{i-j} - \frac{\sin\left(\frac{\pi(L-\delta)(i+j)}{L}\right)}{i+j}\right)|\epsilon_i\rangle\langle\epsilon_j|\delta_{\epsilon_j - \epsilon_i-\hbar\omega}.
    \end{split}
\end{equation}
Considering the fact $\delta$ is an infinitesimally small quantity with $\delta \to 0$, a prefactor of $\frac{1}{\delta^3}$ must be added to the expressions above to ensure that the matrix elements of $\hat{A}_\alpha(\omega)$ remain finite and non-zero.  In the end, the results of $\hat{A}_L(\omega)$ and $\hat{A}_R(\omega)$ are
\begin{equation}
\begin{split}
    \hat{A}_L(\omega) &= \lim_{\delta\to 0} \frac{1}{\delta^3} \frac{1}{\pi} \sum_{i,j}\left(\frac{\sin\left(\frac{\pi\delta(i-j)}{L}\right)}{i-j} - \frac{\sin\left(\frac{\pi\delta(i+j)}{L}\right)}{i+j}\right)|\epsilon_i\rangle\langle\epsilon_j|\delta_{\epsilon_i - \epsilon_j-\hbar\omega}\\
    & = \sum_{i,j}\frac{2\pi^2 i j}{3L^3} |\epsilon_i\rangle\langle\epsilon_j|\delta_{\epsilon_j - \epsilon_i-\hbar\omega}
\end{split}
\end{equation}
and
\begin{equation}
\begin{split}
    \hat{A}_R(\omega) &= \lim_{\delta\to 0} \frac{1}{\delta^3} \frac{1}{\pi} \sum_{i,j}\left(\frac{\sin\left(\frac{\pi(L-\delta)(i-j)}{L}\right)}{i-j} - \frac{\sin\left(\frac{\pi(L-\delta)(i+j)}{L}\right)}{i+j}\right)|\epsilon_i\rangle\langle\epsilon_j|\delta_{\epsilon_i - \epsilon_j-\hbar\omega}\\
    & = \sum_{i,j}\frac{2\pi^2 i j}{3L^3}(-1)^{i+j} |\epsilon_i\rangle\langle\epsilon_j|\delta_{\epsilon_j - \epsilon_i-\hbar\omega},
\end{split}
\end{equation}
respectively.

\subsection*{Details of the relaxation process in the quantum model}\label{sec:AppC}

Here we study the energy relaxation behavior for the quantum bath model at different temperatures. Initially, the system is prepared in a state where all eigenstates up to level $N_{cut}$ are equally populated, i.e. $\rho_{ij} = \delta_{ij}/N_{cut}$. At time $t=0$, it is connected 
to thermal baths on the left and right, both maintained at the same temperature $T_L=T_R=T$. The results of relaxation behaviors are presented in Fig.~\ref{fig:fig6}(a).
In the high-temperature regime, the system behaves similarly to the classical MCMC Maxwell bath model, exhibiting faster relaxation when coupled to a higher-temperature bath.
However, when the bath temperature is well below the minimum energy gap $(\epsilon_2 - \epsilon_1)/k_B$, the system’s relaxation rate becomes temperature-independent. These findings are further supported by the temperature dependence of the second through the sixth eigenvalues of the superoperator $\mathcal{L}$ with the largest real parts, as illustrated in Fig.~\ref{fig:fig6}(b). It is worth noting that, although the relaxation rate does not change in the low-temperature regime, the quantum model is still consistent with the third law of thermodynamics, as it requires an infinite time to reach the exact steady state corresponding to the zero eigenvalue of the superoperator $\mathcal{L}$. 

\begin{figure}
    \centering
    \includegraphics[width=0.6\linewidth]{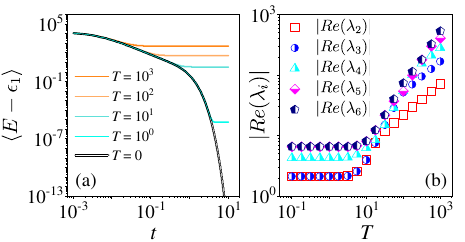}
    \caption{(a) Time dependence of average energy of the particle with different temperature. Here we set $T_L=T_R=T$. The temperature decreases gradually from top to bottom. (b) Temperature dependence of the second, third, fourth, fifth and sixth largest real part of the eigenvalues of the superoperator $\mathcal{L}$.}
    \label{fig:fig6}
\end{figure}

\subsection*{Comparison of quantum results between different length $L$}\label{sec:AppD}

For the classical gas model, both flux and conductivity exhibit a clear $1/L$-dependence on the channel length $L$. This arises from the transport times $t_{L\to R}$ and $t_{R\to L}$  in Eq. \eqref{eq:cl_flux}, which scale linearly with $L$. This scaling behavior is further confirmed by the Maxwell bath results in Eqs. \eqref{eq:maxwellflux} and \eqref{eq:cl_kappa}. 
In the following, we show that by rescaling $\gamma$ as $\gamma L^{-0.5}$, the quantum system exhibits the same $L$-dependence as in the classical case. In Fig. \ref{fig:fig7} (a) we show that upon rescaling $\gamma$ as $\gamma L^{-0.5}$ and temperature as $T/L^2$, the quantum flux $J^q$ for different channel lengths $L$ collapses onto a single curve when further rescaled as $J^qL^4$, consistent with the scaling argument presented in the main text. Fig. \ref{fig:fig7} (b) shows the $1/L$-dependence of the quantum thermal conductivity, after dividing out the temperature contribution. 

\begin{figure}[b]
    \centering
    \includegraphics[width=0.65\linewidth]{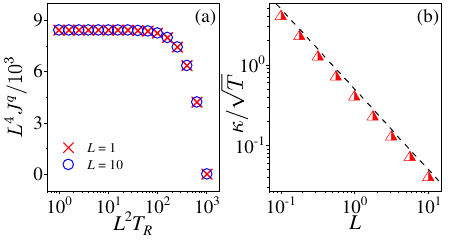}
    \caption{(a) Results of quantum thermal flux for different length $L$. For $L=1$, $T_L=10^3$ and $\gamma = 5.255$; for $L=10$, $T_L=10$ and $\gamma = 1.662$. (b) Length dependence of the quantum thermal conductivity. The coupling strength $\gamma$ is rescaled as $5.255\times L^{-0.5}$. 
    Dashed line is for the reference line $\kappa/\sqrt{T}\propto L^{-1}$.}
    \label{fig:fig7}
\end{figure}

\end{widetext}
\end{document}